\documentclass[preprint,authoryear,12pt]{elsarticle}
\usepackage{graphicx}
\usepackage{amssymb}

\usepackage[%ps2pdf,%
breaklinks=true,%
colorlinks=true,%
pdfauthor={First Author et al.},%
pdftitle={Template for manuscripts in Advances in Space Research}%
]{hyperref}

\journal{Advances in Space Research}
\newcommand{\fix}[1]{{\color{black}#1}} % Problematic parts. Use this way: \fix{what??}
\newcommand{\fixx}[1]{{\color{black}#1}}

\begin{document}

%%%%%%%%%%%%%%%%%%%%%%%%%%%%%%%%%%%%%%%%%%%%%%%%%%%%%%%%%%%%%%%%%%%%%%%%%%%%%
%% Frontmatter
\begin{frontmatter}


% Use the tnoteref command within \title and fnref within \author or \address for footnotes;
% use the corref command within \author for corresponding author footnotes;
% \author{Name\corref{cor1}\fnref{label2}}
% \address{Address\fnref{label3}}

\title{Impact of Rogue Active Regions on Hemispheric Asymmetry}
%\tnotetext[footnote1]{This template can be used for all publications in Advances in Space Research.}

% Use optional labels to link authors explicitly to addresses:

\author{{Melinda Nagy}}
\address{Department of Astronomy, E\"{o}tv\"{o}s Lor\'and University, P\'azm\'any P\'eter s\'et\'any 1/A. H-1117 Budapest, Hungary}
%\cortext[cor]{Corresponding author}
%\fntext[footnote2]{Additional information regarding the corresponding author}
\ead{M.Nagy@astro.elte.hu}

\author{{Alexandre Lemerle }}
\address{Coll\`ege Bois-de-Boulogne, 10555 av. Bois-de-Boulogne, \\ Montr\'eal, QC, H4N 1L4, Canada}
\ead{lemerle@astro.umontreal.ca}

\author{{Paul Charbonneau }}
\address{D\'epartement de Physique, Universit\'e de Montr\'eal, 2900 boul. Edouard-Montpetit, Montr\'eal, QC, H3T 1J4, Canada}
%\cortext[cor]{Corresponding author}
%\fntext[footnote2]{Additional information regarding the corresponding author}
\ead{paulchar@astro.umontreal.ca}

\begin{abstract}
The solar dipole moment at activity minimum is a good predictor of the strength of the subsequent solar cycle. Through
a systematic analysis using a state-of-the-art $2\times2$D solar dynamo model, we found that bipolar magnetic regions (BMR) with atypical characteristics can modify the strength of the next cycle via their impact on the
buildup of the dipole moment as a sunspot cycle unfolds.
In addition to summarizing these results, we present further effects of such ``rogue'' BMRs.
These have the ability to generate hemispheric asymmetry in the subsequent sunspot cycle, since they modify the polar cap flux asymmetry of the ongoing cycle. We found strong correlation between the polar cap flux asymmetry of cycle $i$ and the total pseudo sunspot number asymmetry of cycle $i+1$. Good correlation also appears in the case of the time lag of the hemispheres of cycle $i+1$.
\end{abstract}


\end{frontmatter}

\parindent=0.5 cm

%%%%%%%%%%%%%%%%%%%%%%%%%%%%%%%%%%%%%%%%%%%%%%%%%%%%%%%%%%%%%%%%%%%%%%%%%%%%%
%% Main text
\section{Introduction: Prediction of Hemispheric Asymmetry}

It is well known, starting already with the researches of Rudolf Wolf in the
nineteenth century, that the sunspot cycle is not strictly periodic. Significant
cycle-to-cycle variability is observed in both the amplitude and duration
of the cycle. The hemispheric asymmetry in sunspot coverage
was already noticed by \cite{Waldmeier1955}, who also pointed out that such
asymmetry could be sustained for several years.
\cite{Babcock1959} reported that this asymmetry appears in the case of the polar field reversal as well.
Based on observational data for cycles 12-23, \cite{Norton2010} analyzed the asymmetry and proposed 20\% as upper limit of sunspot area asymmetry. Regarding the phase lag asymmetry for the same period of time, \cite{McIntosh2013} found a roughly four cycles long periodicity. \cite{Zolotova2010} investigated the phase difference of sunspot cycles in the hemispheres for a longer duration, back to the Maunder Minimum and found secular variation. On the other hand, this periodicity appears only in the sign of the phase lag but not in its magnitude \citep{Norton2014}. %For a detailed review of observational characteristics of solar hemispheric asymmetry see \cite{Norton2014}.
\cite{Hathaway2016} predicted the hemispheric asymmetry of cycle 25 by extrapolating the polar fields of cycle 24 using their Advective Flux Transport model.
According to their results, the Southern hemisphere should dominate the North.
However, from a purely statistical point of view, the available polar field data (eg. \citealt{Munoz2012}) is insufficient to infer a significant
correlation from past cycles.

\cite{Belucz2013} investigated the hemispheric asymmetry generated
in a 2D flux transport dynamo model by inserting a second meridional circulation cell on the southern hemisphere with different amplitude, latitude and depth. They found significant hemispheric asymmetry depending the properties of the second cell. Using the same model, \cite{Shetye2015} focused on the effects of the meridional inflow towards the activity belts. According to their results, the intense inflow in one hemisphere leads to stronger toroidal fields and this asymmetry is sustained for more than one cycle.

\cite{Karak2017}, using their 3D surface flux transport and Babcock-Leighton solar dynamo model (\citealt{Miesch2014}; \citealt{Miesch2016}),
investigated the influence of the tilt angle distribution by adding random scatter on Joy's law and a tilt-angle saturation was also added to the model (on this point
see also \citealt{Lemerle2017}).
The SpotMaker algorithm they use places new BMRs in each hemispheres, with a set time
delay in order to avoid artificially imposing hemispheric symmetry. One of their results is the hemispheric asymmetry appearing in the polar fluxes
was only weakly correlated to the toroidal flux of the subsequent cycle. Due to the
strong diffusive coupling between the hemispheres in the model, the asymmetry reduces quickly.

In the present paper \fix{we extend the work of \citep{Nagy2017} on
the hemispheric asymmetry triggered by rogue BMRs. We present a detailed analysis of how rogue BMRs affect the hemispheric asymmetry of the polar cap flux, including prediction of the asymmetry level of the subsequent cycle based on the polar field asymmetry. We also investigate whether the asymmetry in the model shows periodicity or temporal persistence in its characterizing parameters.
Following \citep{Nagy2017}, our analysis is based on simulations carried out
using the recent $2\times2$D dynamo model of \citealt{Lemerle2015, Lemerle2017}.}

\section{Rogue BMRs in the $2\times2$D Dynamo Model}

The \cite{Lemerle2017} solar cycle model invokes differential rotation shear
and the regeneration of the solar dipole via surface decay of active regions
(the so-called Babcock-Leighton mechanism) as its primary inductive mechanisms.
This mean-field-like kinematic model
couples a 2D surface flux transport module
(SFT) with a
2D axisymmetric flux transport dynamo (FTD). The SFT component provides the azimutally averaged radial field
serving as the upper boundary condition for the FTD
simulation, while the FTD module drives the SFT through the emergence
of new bipolar magnetic region (BMR). This step is based
on a semi-empirical emergence function that sets the
probability of a BMR emerging (radially)
as a function of toroidal magnetic field $B_t$ at
the bottom of the convective zone in the FTD module.
\fix{Motivated by the modelling of the destabilization and buoyant rise of thin
magnetic flux ropes initially located immediately beneath the base of the
solar convection zone \citep[see, e.g.,][and references therein]{Fan2009},
a lower threshold
on $B_t$ is introduced, below which the emergence probability
vanishes. The presence of a threshold implies that the dynamo is not self-excited,
as the internal magnetic field must remain above threshold for regeneration of
the surface dipole to take place. Above this threshold, the proportionality constant
$K$ between emergence probability and
$B_t^{1.5}$ acts as the dynamo number for the model, since it effectively
sets the mean surface dipole growth rate for a given internal toroidal
field strength.}
Properties of the
new BMR --- emergence latitude, longitude, flux, angular separation, tilt --- are randomly drawn
from distribution functions for these quantities built from observed
statistics of solar active regions during solar cycle 21, as described in Appendix A of
\citealt{Lemerle2015}.

Because the model is kinematic and includes a steady quadrupole-like meridional
flow in the FTD module, the only physical mechanism available to couple
the two hemispheres is diffusive transport, operating in both the SFT and FTD modules.

The only amplitude-limiting nonlinearity included in the model
is a
reduction of the average BMR tilt angle $\alpha$ as a function of the deep-seated
toroidal field strength $B_t$, parametrized
according to the following ad hoc algebraic formula:
\begin{equation}\label{eq:tiltquenching}
    \alpha_q = \frac{\alpha(\theta)}{1 + (B_t/B_q)^2},
\end{equation}
where $B_q$ is the quenching field amplitude, and $\alpha$ is the reference
tilt variation with latitude, i.e., Joy's law \citep{McClintock2013}.
\fix{Lacking any reliable information on the manner in which the emerging flux ropes
producing BMRs disconnect from the underlying toroidal magnetic flux system,
we do not implement any flux reduction in the FTD module when emergences
are introduced in the SFT module.}

The main advantages of the $2\times2$D model are its high numerical efficiency
and the fact that it is calibrated to follow accurately the
statistical properties of the real Sun. The complete
latitude--longitude representation of the simulated solar surface in
the SFT component further makes it possible to achieve high spatial
resolution and account for the effect of individual active region
emergences.

The reference solar cycle solution presented in \citet{Lemerle2017} is
defined by 11 adjustable parameters, for which optimal values were obtained
via a formal optimization by a
genetic algorithm. The algorithm was designed to minimize the differences between the
spatiotemporal distribution of emergences produced by the model, and
the observed sunspot butterfly diagram during cycle 21.

\begin{figure}
  \centering
  \includegraphics[width=\textwidth]{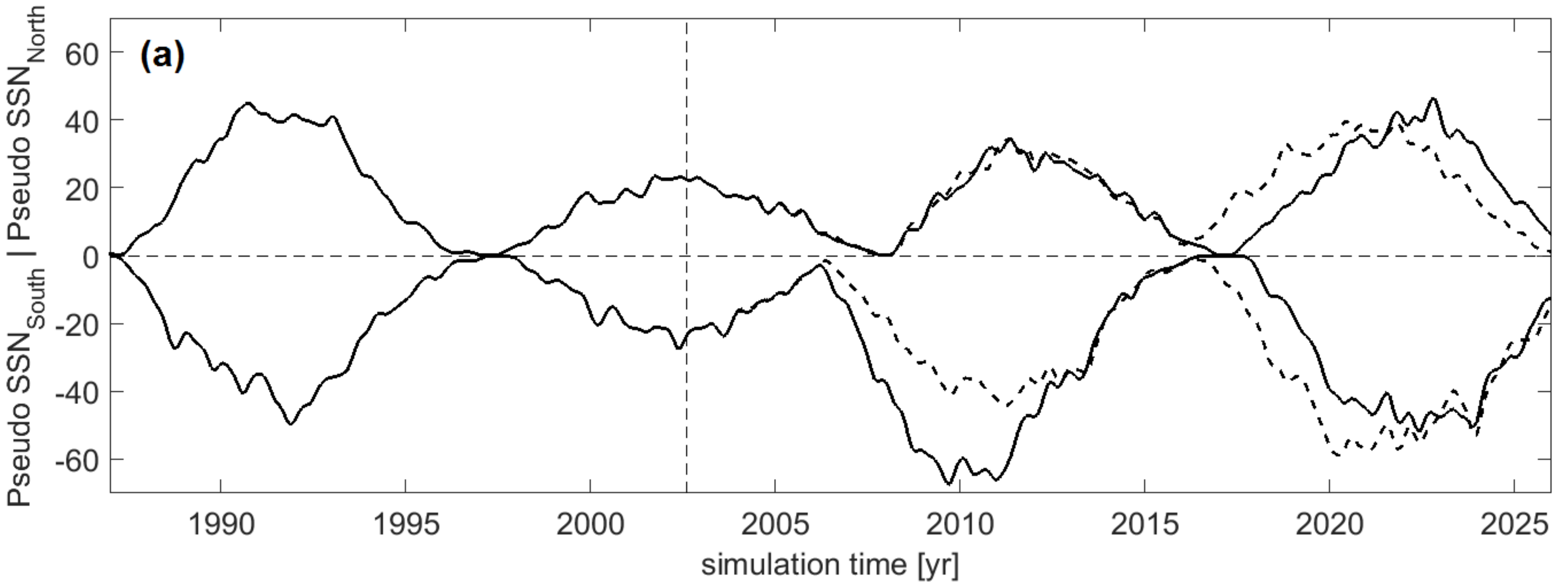}\\
  \includegraphics[width=\textwidth]{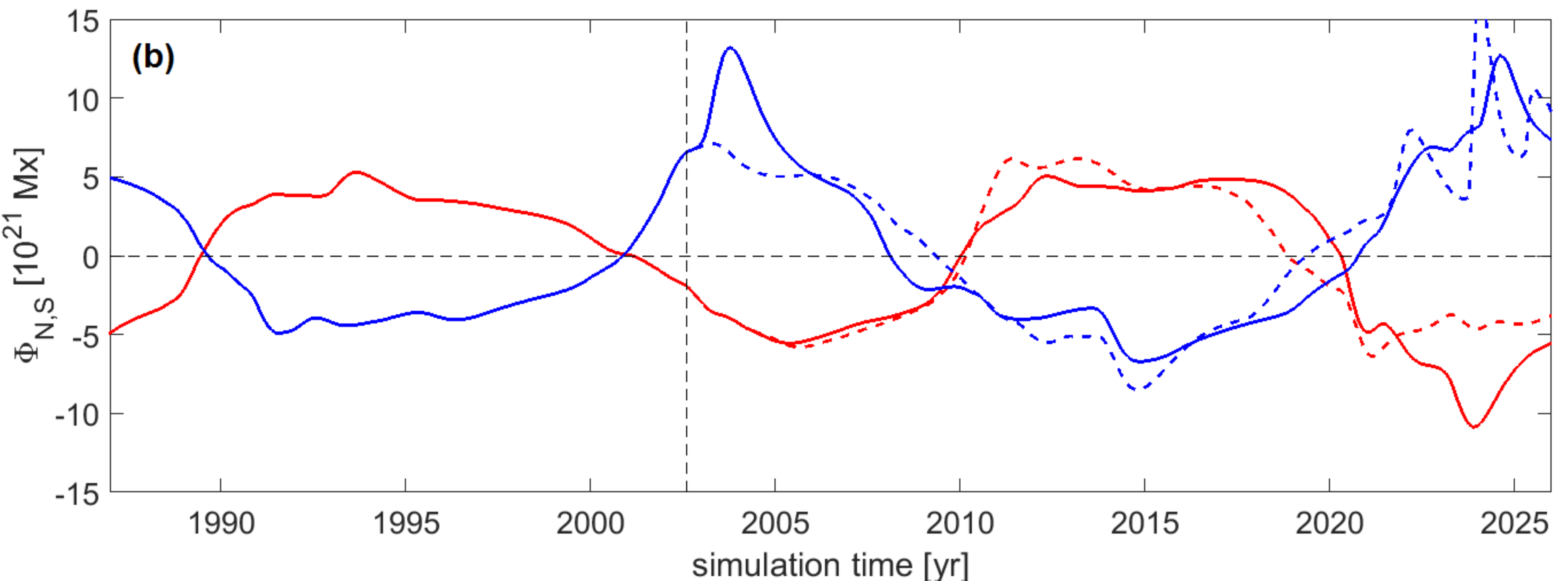}\\
  \includegraphics[width=\textwidth]{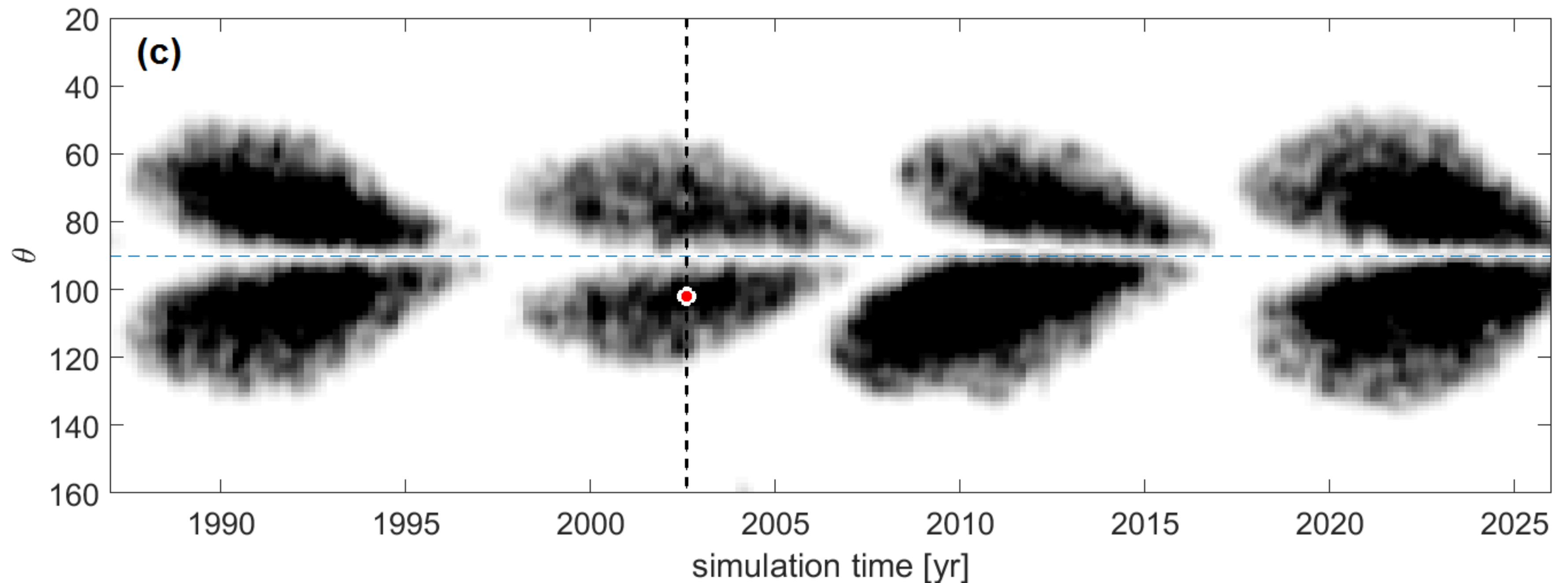}\\
  \caption{A short segment of the reference dynamo solution used for the analyses
discussed in the present paper.
Panel (a) shows the pseudo sunspot number time series separately for the hemispheres.
The solid lines shows the reference simulation run.
On panel (b) hemispheric time series of the polar cap flux are plotted, in
red and blue for the Northern and Southern hemispheres, respectively.
On both of these panels dashed lines
pertain to a experiment
in which a single large ``rogue'' BMR was removed from the simulation at the time indicated
by the vertical black dashed line.
Panel (c) shows the pseudo-sunspot butterfly diagram of the reference simulation plotted
as solid lines in panels (a) and (b).
The gray scale encodes the density of emerging BMRs, and
the red dot indicates the position of the rogue BMR removed from the simulation
to yield the dashed time series in (a) and (b).
}\label{fig:refsolution}
\end{figure}

Figure \ref{fig:refsolution} shows a portion of the reference dynamo solutions
used for the analyses presented in what follows. The solid lines on the
top and middle panels
show time series of hemispheric pseudo-sunspot number and polar cap flux, \fix{which is calculated here as the surface integral of the radial magnetic field over a latitudinal extent of $20^{\circ}$ from the poles.}
The bottom panel shows the corresponding time-latitude ``butterfly''
diagram for the spatial density of emerging BMRs, encoded as a grayscale.
The red dot indicates the time and latitude at which a large ``rogue''
BMR emerged in this simulation,
its properties being listed in the first row of Table \ref{tab:BMRs}.
Artificially removing this single BMR from the simulation leads to
a markedly different subsequent evolution of the dynamo solution,
as shown by the dashed time series on panels (a) and (b)
of Figure \ref{fig:refsolution}.

The reference solution plotted on Figure \ref{fig:refsolution}
is the same as adopted in the numerical experiments of
\citet{Nagy2017} where the impact of individual ``rogue'' BMRs were analyzed. These peculiar active regions were identified based on their contribution to the global dipole moment defined as follows:
\begin{equation}
 \delta D_{\mathrm{BMR}} \approx F \,d \, \sin\alpha \,  \sin\theta,
 \label{eq:thenumber}
\end{equation}
where $F$ is magnetic flux, $d$ is the
angular separation of the two polarities, $\alpha$ is the tilt angle, $\theta$
is the colatitude. According to this expression, BMRs with high flux content, tilt angle and angular separation, close to the equator influence the most the building up dipole moment, and therefore the strength of the subsequent cycle \fix{as suggested by \citet{Jiang2015} as an explanation for the low amplitude of Cycle 24}.

\begin{figure}[t!]
  \centering
  \begin{minipage}{0.55\textwidth}
  \includegraphics[width=0.95\linewidth]{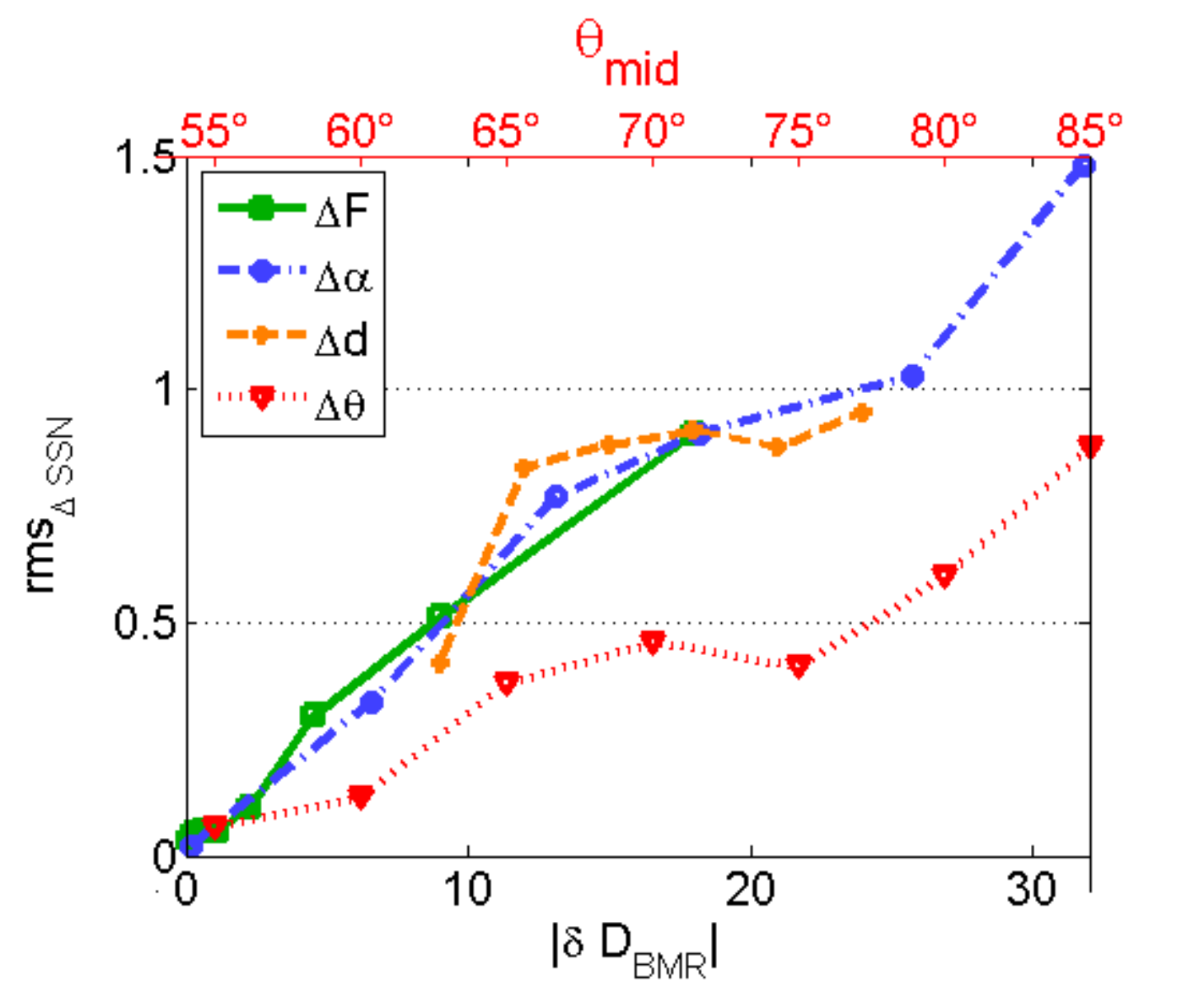}
  \end{minipage}
  \begin{minipage}{0.44\textwidth}
  \caption{Average effect of varying the properties of a BMR (2nd data
row of Table \ref{tab:BMRs}), inserted in the simulations at cycle
maximum, on the amplitude of the subsequent cycle. Variations in BMR
flux (green), tilt angle (blue) and  polarity separation (orange) are
converted to the contribution to the dipole moment according to
Equation (\ref{eq:thenumber}), while the varying colatitudes (red) are
shown on the top axis (adapted from Figure 11 in \citealt{Nagy2017}).}\label{fig:summarize}
  \end{minipage}
\end{figure}

\citet{Nagy2017} carried out
several numerical experiments aiming to study how the parameters of BMRs affect the next, or even the ongoing cycle. A selected ``test'' BMR
with specified properties ($F$, $d$, etc.)
was inserted manually into simulations while the parameters of the active region were varied separately during each experimental series. The test-BMR emerged spontaneously during the reference simulation with parameters listed in the second row of Table \ref{tab:BMRs}. The experiments were performed for three cycles with average, below average and above average amplitudes, respectively.
In each case two series of experiments were carried out with Hale (anti-Hale) test-BMR in order to increase (decrease) the dipole moment of the examined cycle. The characteristics of the
test-BMR -- emergence time and latitude, flux, tilt angle and angular separation -- were changed one by one in order to map the impact of each property on the subsequent simulated cycle.
The results of these experiments are summarized in Figure \ref{fig:summarize}.
By jointly varying the flux, tilt angle or separation of the test-BMR, similar results were obtained for the amplitude of the upcoming cycle. This is because
these quantities have a combined effect in the form of Equation (\ref{eq:thenumber}). The impact of the emergence latitude is also shown in Figure \ref{fig:summarize} (red curve, top axis).
The effect of the BMR decreases as a function of the emergence latitude, but is still significant $20^{\circ}$ away from the Equator. The emergence epoch is also an important factor; the strongest impact on the subsequent cycle is expected when the test-BMR appears at cycle maximum,
and diminishes gradually as the rogue BMR is forced to emerge later and later in the descending
phase of the cycle.
When the emergence occurs during the rising phase of the perturbed cycle, it modifies the ongoing cycle as well.

\begin{table}[t!]
    \center
  \caption{Parameters of active regions discussed in the
paper.  Colatitudes $\theta_{\mathrm{lead}}$ and
$\theta_{\mathrm{trail}}$ are the
  latitudinal positions of leading and trailing polarities;  $F$ is
the flux of the trailing polarity ($F_{\mathrm{trail}} =
-F_{\mathrm{lead}}$); $\alpha$ is the tilt angle and $d$ is the
angular separation of leading and trailing polarities. $\delta
D_{\mathrm{BMR}}$, the contribution of the BMR to the global dipole
moment, is defined according to Equation (\ref{eq:thenumber}). J/H
indicates whether the active region is (anti-)Joy/(anti-)Hale. In the
case of the second row a J/H (J/a-H) test-BMR increases (decreases)
the dipole moment during the experiments detailed in Section 5 of
\citep{Nagy2017}. }\label{tab:BMRs}
  \begin{tabular}{ c  c  c  c  c  c  c }
  \hline
  $\theta_{\mathrm{lead}}$ & $\theta_{\mathrm{trail}}$ & $F $ [$10^{23}$ Mx] & $\alpha$  & $d$ & $\delta D_{\mathrm{BMR}}$[$10^{23}$ Mx] & J/H  \\
  \hline
  112$^{\circ}$           & 118.6$^{\circ}$           & 4.39              & $-11.08^{\circ}$ & 34.08$^{\circ}$   & $-0.5124$ &  J/H\\
  89.5$^{\circ}$           & 82.1$^{\circ}$            & --1.39              & 13.98$^{\circ}$  & 30.97$^{\circ}$   & --0.1810 & J/H  \\
                           &                           &                     &                  &                   &          & J/a-H \\
  \hline
  \end{tabular}

\end{table}

\section{Hemispheric Asymmetry due to Rogue BMRs}

As proposed by \cite{Hathaway2016} in the context of variation in the surface
meridional flow, the hemispheric asymmetry of a solar cycle originates from the polar cap flux asymmetry during the preceding cycle. \citet{Nagy2017} analyzed whether the polar cap flux asymmetry is a good indicator of the upcoming simulated cycle's asymmetry in the $2\times2$D model.

\begin{figure}
  \centering
  \includegraphics[width=\textwidth]{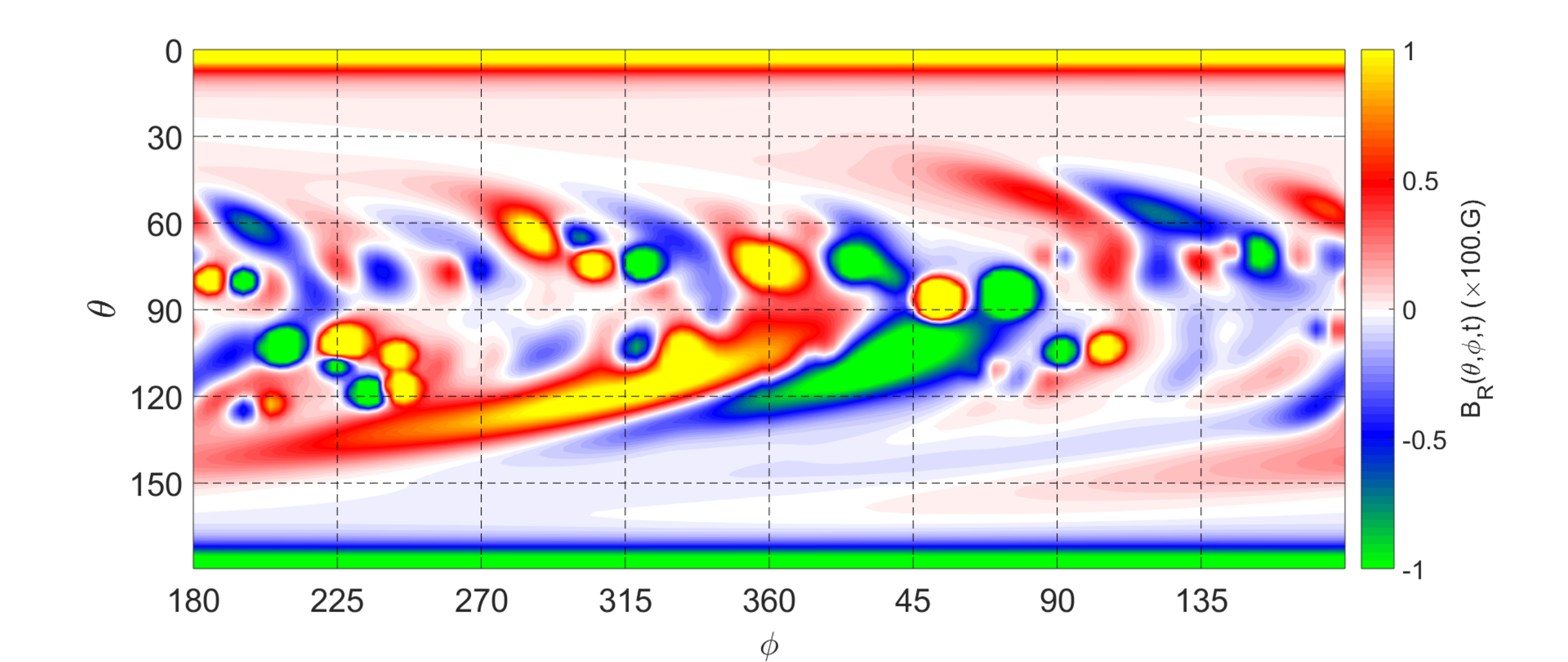}
  \includegraphics[width=\textwidth]{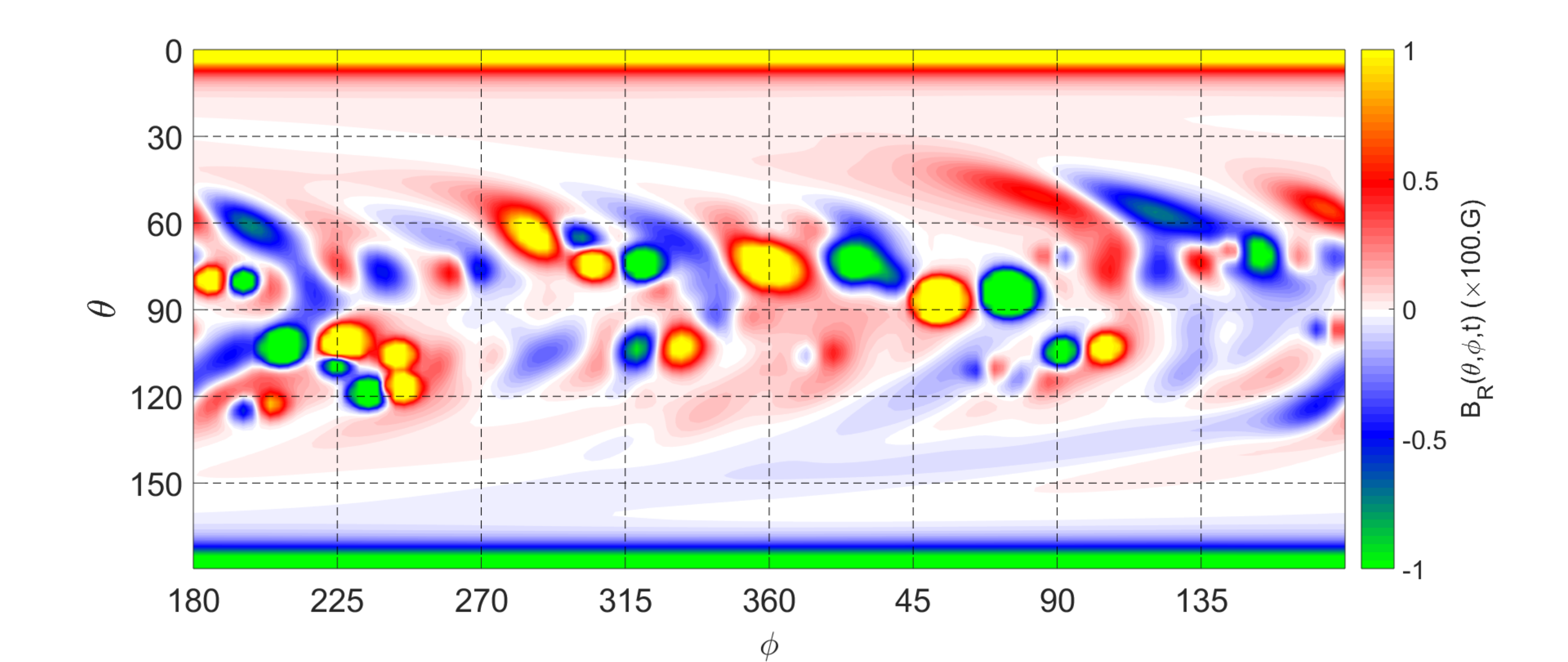}
  \caption{\fixx{On the top we plot the synoptic magnetogram of the rogue BMR listed in the first row of Table \ref{tab:BMRs} at central longitude $\phi\sim360^{\circ}$ and simulation time $t = 1992.72$, six months after the time of emergence (see the corresponding time series in Figure \ref{fig:refsolution} and \ref{fig:magnetogr}).
The bottom panel shows the magnetic field distribution without the emergence of this active region at the same epoch.  Here (and on Fig.~\ref{fig:magnetogr} following), the color scale is strongly saturated to make the weaker magnetic fields visible.
}}\label{fig:synopticRogue}
\end{figure}

Comparison of the solid and dashed lines on Figure \ref{fig:refsolution}(a) and (b) indicates that a single, large BMR with unusual characteristics
can have a large and persistent impact on
hemispheric asymmetry in the resulting dynamo solution.
\fixx{The top panel on Figure \ref{fig:synopticRogue} shows a synoptic magnetogram extracted from a simulation in which the rogue BMR listed in the first row of Table \ref{tab:BMRs} was inserted at simulation time $t = 1992.72$, one cycle before the strongly asymmetric cycle that we will discuss in the next section.
This snapshot is extracted six months after emergence of the rogue BMR.
This
is an anti-Hale BMR, with polarity ordering opposite to that which should
characterize its hemisphere, so that the emergence impedes the build up
of the Southern polar fields. The poleward surge of positive polarity (red)
can be seen quite clearly.
For comparison, the bottom
panel shows a second synoptic magnetogram, extracted at the same time in
a parent simulation without artificial insertion of the rogue BMR.
Comparing the two snapshots reveals how
the positive trailing flux of the decaying rogue has strongly
altered
the pattern of magnetic flux transport to the Southern polar regions.
This pattern is qualitatively similar to that highlighted by
\citet{Upton2018}, with the large
active region AR12192
emerging in October 2014 and producing a poleward stream of positive
magnetic polarity which weakened the buildup of the southern polar cap
negative magnetic field.
}

\begin{figure}
  \centering
  \includegraphics[width=\textwidth]{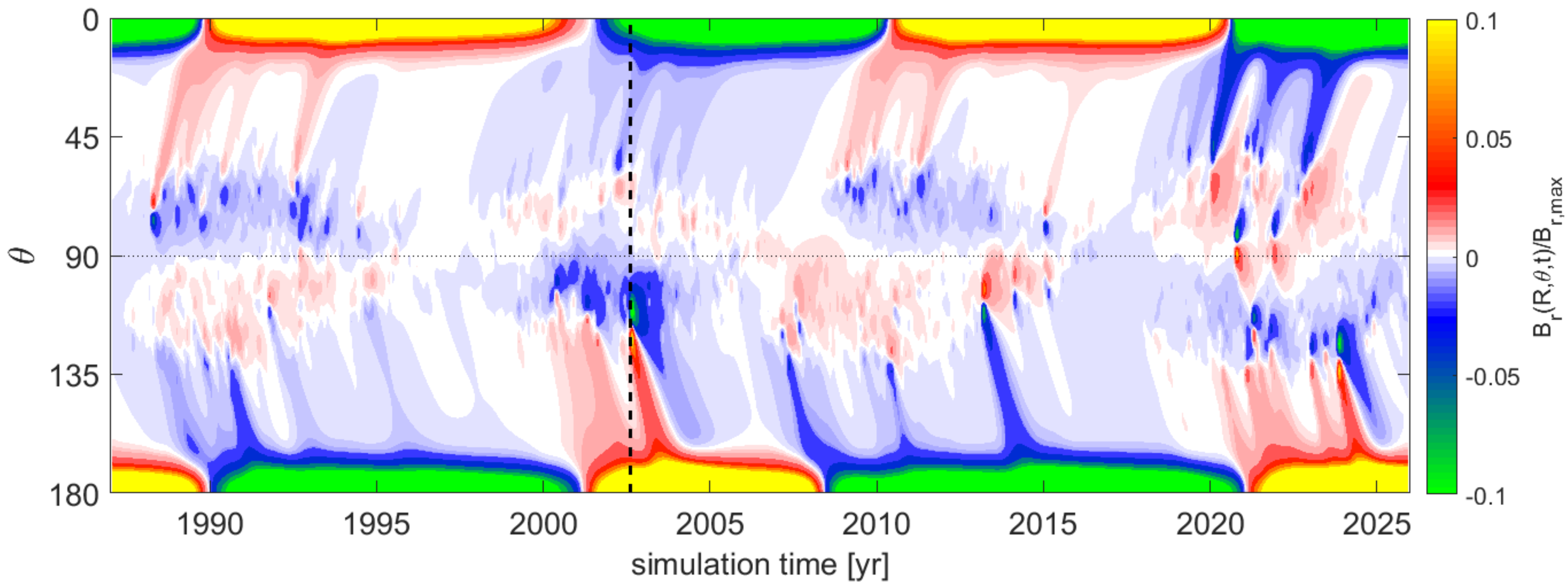}\\
  \includegraphics[width=\textwidth]{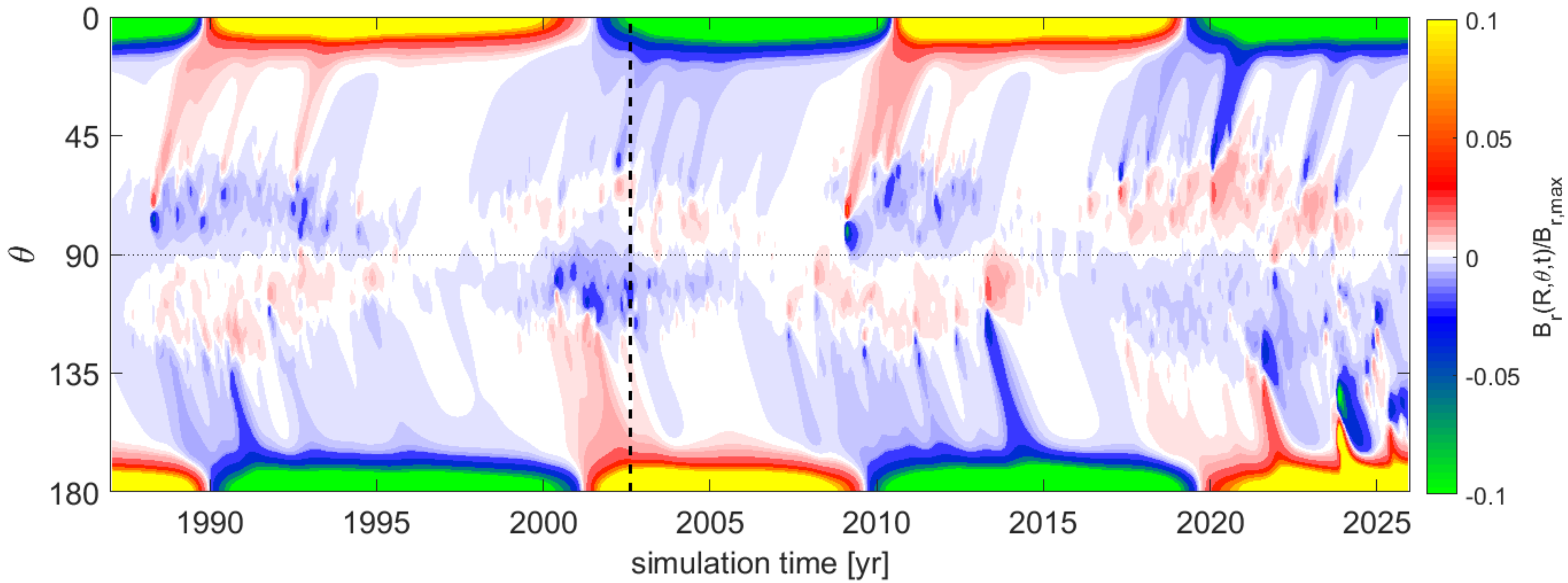}\\
  \caption{The top panel shows a synoptic magnetogram of the surface radial magnetic
field component in the reference solution of Figure \ref{fig:refsolution}. The bottom panel
shows the synoptic magnetogram resulting from removal of the rogue BMR at the
time indicated by the vertical dashed line, corresponding to the dashed time
series on panels (a) and (b) of Figure \ref{fig:refsolution}.}\label{fig:magnetogr}
\end{figure}

Figure \ref{fig:magnetogr}
offers another example of this effect, for a different rogue event
and now in the form of time-latitude maps
of the zonally-averaged
surface radial magnetic field component. The top panel corresponds to the
reference solution, while the bottom panel shows the magnetogram resulting from the
artificial removal of a single large BMR at the time indicated by the vertical dashed line.
Note in particular how the reversal of the polar field occurs almost
simultaneously in both hemispheres when the rogue BMR is removed, while in the
original reference solution the southern polar cap reverses polarity some two years
prior to the northern hemisphere.

\fix{Note that here the polar cap flux peaks 2--3 years prior to SSN minimum,
while in
the case of the sun this peak usually occurs somewhat closer to cycle minimum.
However, in our model as in the sun,
the peak polar cap flux does turn out
to be a good predictor of the SSN amplitude of the subsequent cycle, so we retain
it as a measure of cycle dipole strenth in all analyses that follow.}

To quantify the level of asymmetry, the normalized asymmetry of the peak polar cap flux ($\Delta_{\Phi}$) produced during the cycles is
compared to two asymmetry measures characterizing the subsequent cycles. These measures are the asymmetry of the total number of emergences in each hemisphere ($\Delta_{\mathrm{SSN}}$), and the time delay between the epochs when the new cycle BMRs first
start to emerge in the North and the South ($\Delta_{T}$).

The asymmetry of the polar cap flux at a given cycle is defined as follows:
        \begin{equation}
        \Delta_{\Phi} = \frac{|\Phi_{\mathrm{N},\mathrm{max}}| - |\Phi_{\mathrm{S},\mathrm{max}}|} {( |\Phi_{\mathrm{N},\mathrm{max}}| + |\Phi_{\mathrm{S},\mathrm{max}}| )/2},
        \end{equation}
    \noindent where $\Phi_{\mathrm{N},\mathrm{max}}$ ($\Phi_{\mathrm{S},\mathrm{max}}$) is the northern (southern) polar cap flux maximum.
The asymmetry of the activity level is defined similarly, but in
terms of the pseudo-sunspot number constructed from the model output:
        \begin{equation}
        \Delta_{\mathrm{SSN}} = \frac{\Sigma \mathrm{SSN}_{\mathrm{N}} - \Sigma \mathrm{SSN}_{\mathrm{S}} } {( \Sigma \mathrm{SSN}_{\mathrm{N}} + \Sigma \mathrm{SSN}_{\mathrm{S}} )/2},
        \end{equation}
    \noindent where $\Sigma \mathrm{SSN}_{\mathrm{N}}$ ($\Sigma \mathrm{SSN}_{\mathrm{S}}$) is the total number of emergences in the northern (southern) hemisphere.
Finally, the time lag between the hemispheres is defined as:
        \begin{equation}
          \Delta_{T} = \frac{t_{\mathrm{N}} - t_{\mathrm{S}}} {( T_{\mathrm{N}} + T_{\mathrm{S}} )/2},
        \end{equation}
    \noindent where $t_{\mathrm{N}}$ ($t_{\mathrm{S}}$) is the beginning epoch of the cycle, while $T_{\mathrm{N}}$ ($T_{\mathrm{S}}$) is the duration of the cycle on the North (South).

\begin{figure}[t!]
      \centering
      \begin{minipage}[b]{1\textwidth}
          \includegraphics[width=0.5\textwidth]{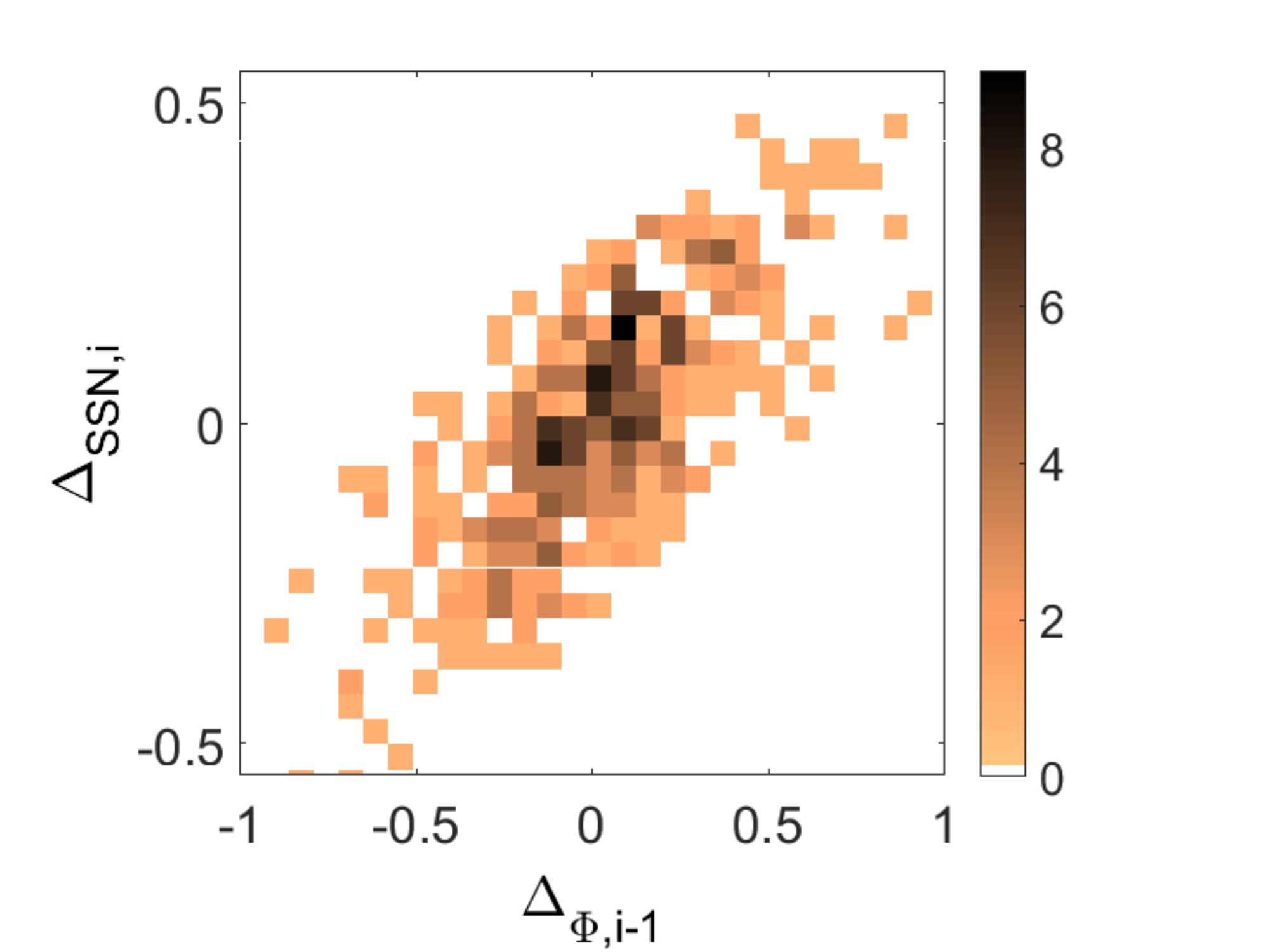}
          \includegraphics[width=0.5\textwidth]{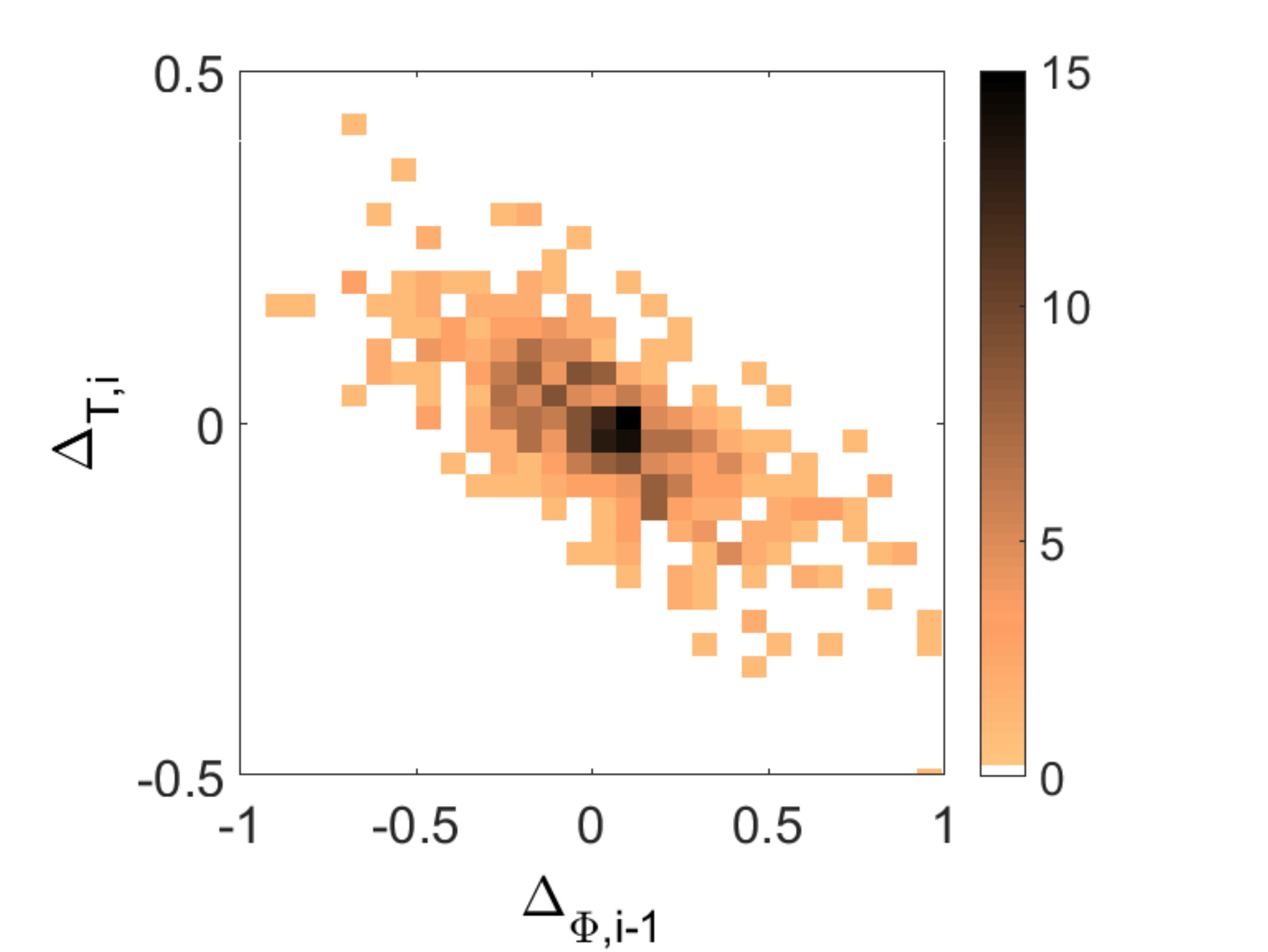}
      \end{minipage}
      \begin{minipage}[b]{1\textwidth}
        \caption{\small{ Two-dimensional histograms of the asymmetry of the hemispheric total
     pseudo-SSN (\textbf{left}) and the time lag between North and South (\textbf{right})
     in pseudo-solar cycle $i$ against the polar cap flux asymmetry during the
     previous cycle for 540 simulated cycles. Some outlier data have been removed.
     The number of cases (cycles) in each bin are
     indicated by the colour codes. The correlation coefficient is
     $r_{\Delta SSN} = 0.7430$ and $r_{\Delta T} = -0.7174$, respectively (adapted from Figure 6 in \citealt{Nagy2017}).
     }}\label{fig:asymmetry500cycles}
      \end{minipage}
    \end{figure}

Upon calculating these asymmetry measures for 540 simulated cycles, \citet{Nagy2017} found strong anticorrelations between the polar cap flux asymmetry of cycle $i$ and time delay, $\Delta_{\mathrm{T}}$ ($r = -0.7174$)
during cycle $i+1$.
In the case of asymmetry in number of emergences of cycle $i+1$, $\Delta_{\mathrm{SSN}}$  the correlation coefficient is $r = 0.7430$ as it is shown in Figure \ref{fig:asymmetry500cycles}. This result shows that in the model the asymmetry of a cycle can be predicted via the asymmetry of the polar cap flux built up during the previous cycle.

\begin{figure}[t!]
  \centering
  \begin{minipage}{\textwidth}
  \centering
  \includegraphics[width=0.4\linewidth]{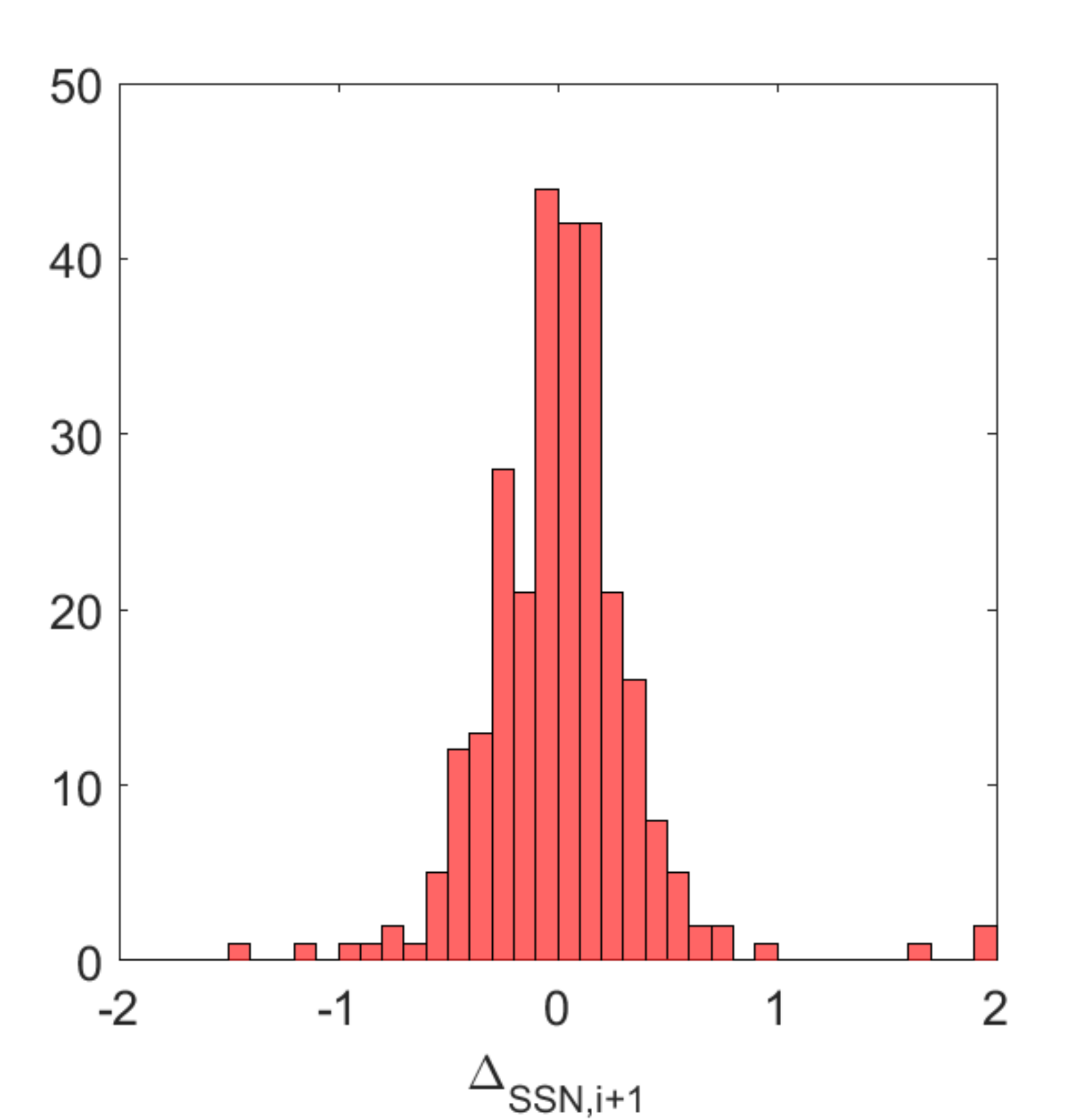}
  \includegraphics[width=0.4\linewidth]{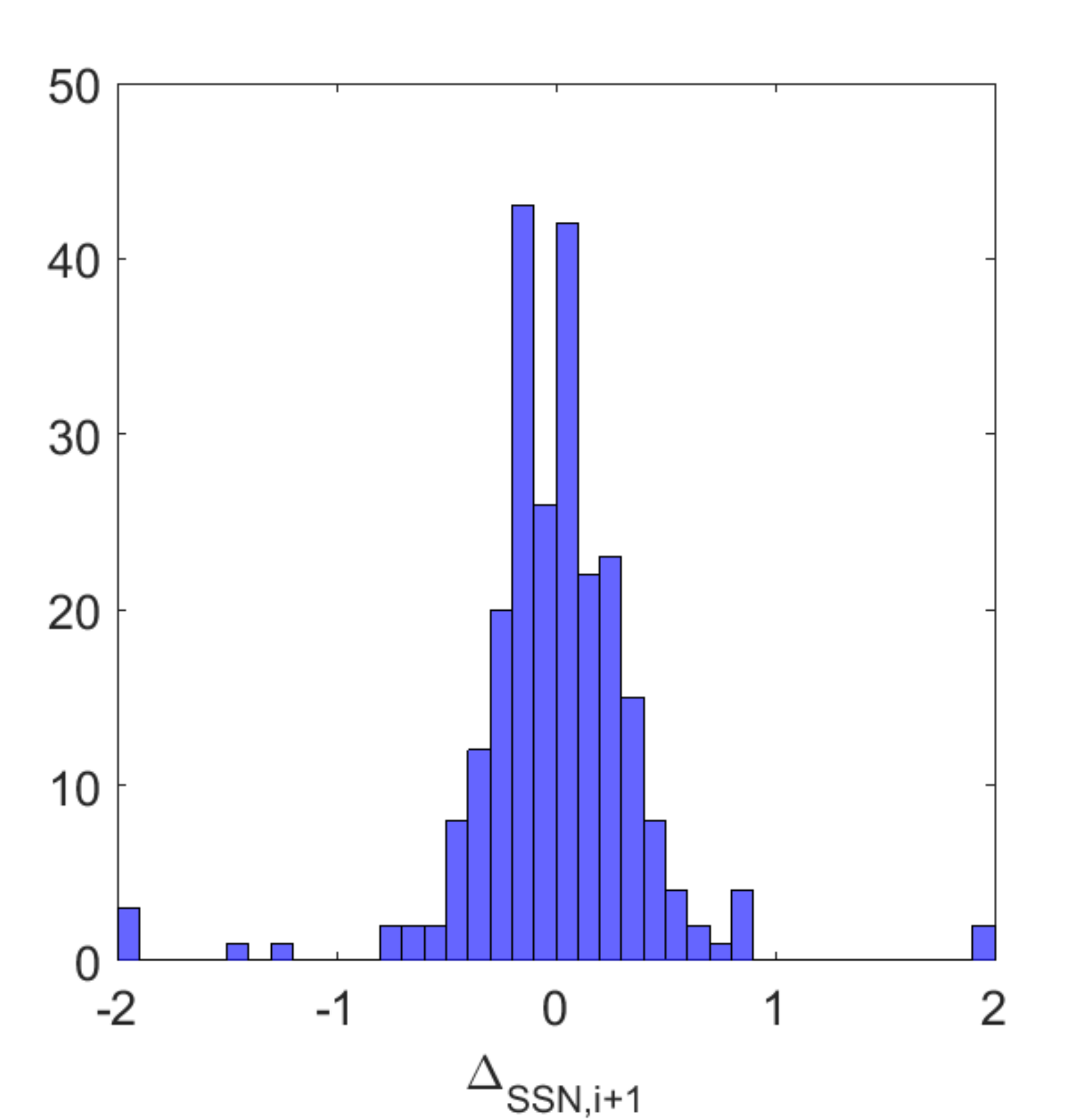}
  \end{minipage}
  \begin{minipage}{\textwidth}
  \caption{ Strength asymmetry histograms of cycles following a North (South) dominated cycle on the left (right) panel. The histograms show the distribution of asymmetry
parameters for all cycles following a Northern-dominated cycle (left) or
Southern-dominated cycle (right). The distributions are approximately Gaussian,
with means and standard deviations are respectively
$0.012$ and $0.36$ in the left panel, and $-0.008$ and $0.41$ for the right panel.
This indicates that hemispheric dominance shows no significant
persistence from one cycle to the next, at least in the (optimal) parameter regime
of this simulation run.
}\label{fig:PeriodHist}
  \end{minipage}
\end{figure}

In order to assess the persistence of hemispheric asymetry, we separate the
simulated cycles in two groups according to hemispheric dominance, as measured
by the quantity $\Delta_{\rm SSN}$ introduced above. For each group, we then
construct the histograms of $\Delta_{\rm SSN}$ values characterising the
cycle following each member of the groups. The resulting histograms are plotted
in Figure
\ref{fig:PeriodHist}. Both are very well fit with Gaussian centered on
$\Delta_{\rm SSN}=0$, with deviation from zero at the $10^{-2}$ level
and standard deviation $\sim 0.4$. This indicates that the probability of
finding a given hemispheric dominance in cycle $n+1$ is independent of
hemispheric dominance in cycle $n$, and thus that hemispheric dominance is
determined by processes operating on inter-cycle timescales.
We repeated this exercise, this time constructing the distribution of asymetry
parameters two cycles in the future instead of one, in order to possibly detect
persistence of hemispheric asymmetry associated with one magnetic polarity
dominating over the other for a few subsequent cycles, a features sometimes
observed in the \cite{Lemerle2017} dynamo solutions. Once again the mean
of the distribution are very close to zero, with standard deviations $\sim 0.4$,
indicating that hemispheric asymmetry does not persist beyond
one cycle in this model.

\begin{figure}[t!]
  \centering
  \includegraphics[width=\textwidth]{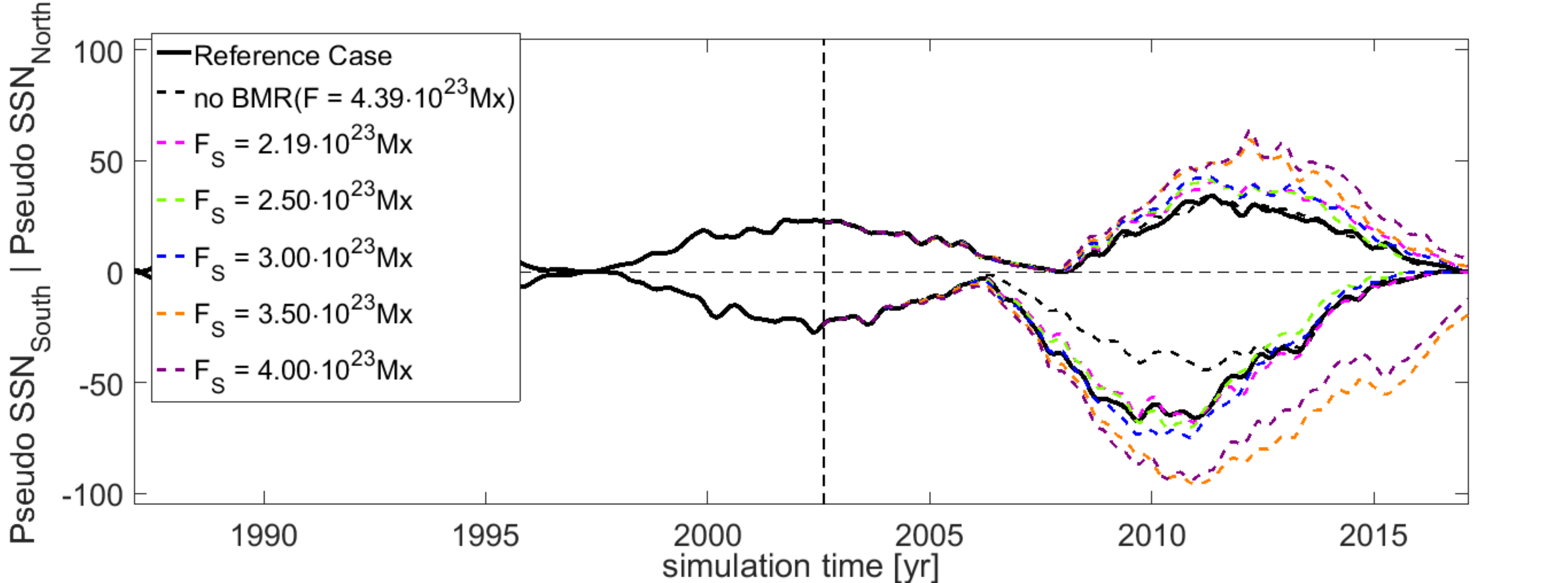}\\
  \includegraphics[width=\textwidth]{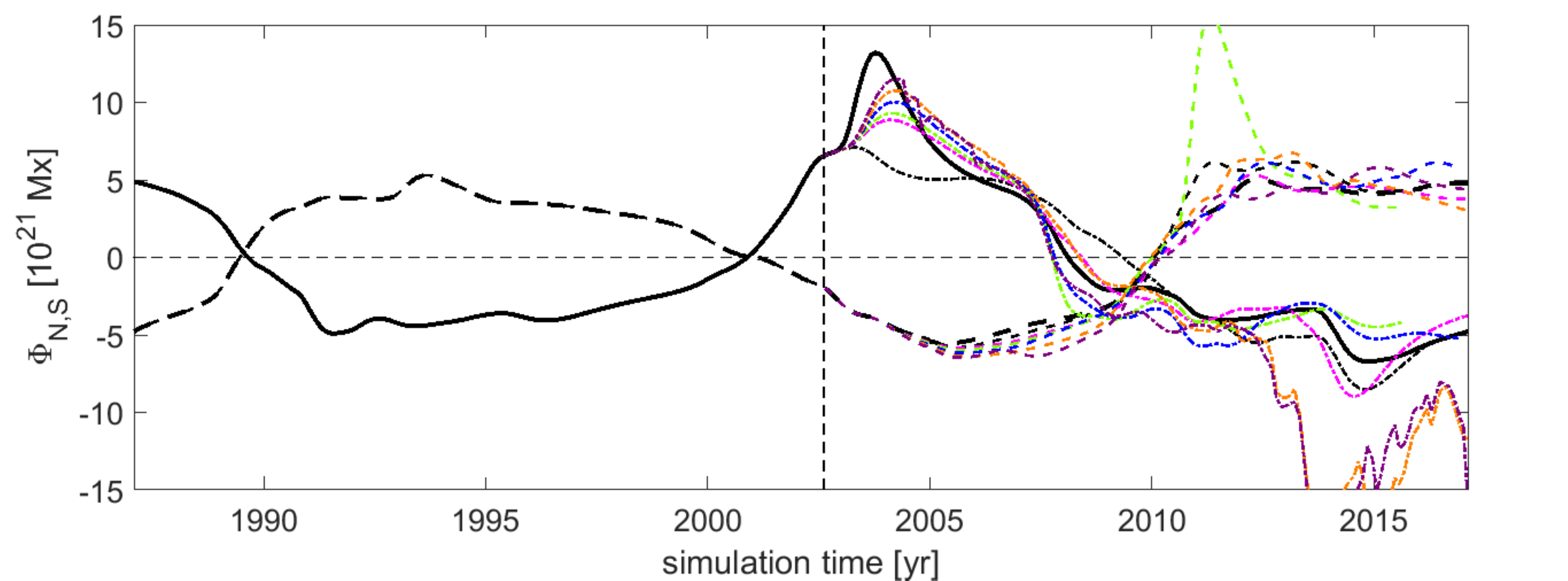}\\
  \caption{The \textbf{top} panel shows how a test-BMR can modify the amplitude of the subsequent cycle, separately in each hemisphere. The properties of this region are listed in the first row in Table \ref{tab:BMRs}. %, displaced to emerge $15^{\circ}$ from the equator.
  Black solid curves indicate the reference solution, with a ``rogue'' BMR emerging
$25^{\circ}$ away from the Equator, at the time indicated by the vertical dashed line.
The black dashed curves refer to a
modified simulation in which this ``rogue'' BMR is removed. Colored curves indicate the results with test-BMRs with different flux, $15^{\circ}$ far from the Equator, as labeled.
On the \textbf{bottom} panel the polar cap flux is shown. Solid and dot-dashed lines indicates the southern polar cap flux while dashed lines correspond to the northern polar cap flux.
}\label{fig:casestudy}%$\psi = 0.5925$
\end{figure}

Based on their numerical experiments, \citet{Nagy2017} identified another interesting effect triggered by the emergence of a rogue BMRs. After such emergences in one cycle, the next cycle tends to be strongly asymmetric. This phenomenon was analyzed using a new test-BMR described in the first line of Table \ref{tab:BMRs}, emerged on the southern hemisphere at cycle maximum, indicated by the vertical dashed line in Figure \ref{fig:casestudy}. According to the previous results, the AR's impact on the upcoming cycle is the strongest at this epoch. On the other hand, the original position of the BMR is a bit far from the Equator.
For this reason, during the experimental runs the test region was replaced to emerge closer, about $15^{\circ}$ far from the equator, within the region where significant effect was observed during the first experimental series.
At this position the active region's flux was decreased from about $4\cdot10^{23}$ Mx down to $2.19\cdot10^{23}$ Mx.
The black solid line in the top panel of Figure \ref{fig:casestudy} shows the reference case when the BMR emerged at the original position, while black dashed line shows the case when the BMR was removed from the simulation. The coloured dashed lines show how the asymmetry changed for various values of the test-BMR's flux, as color-coded.
One can see that the asymmetry is changing according to the flux of the test region. There are slight changes in the amplitude of the northern hemisphere as well due to the
diffusive hemispheric coupling in the model.
The bottom panel of Figure \ref{fig:casestudy} shows that the hemispheric asymmetry already appears in the form of polar cap flux asymmetry during the cycle within which
the test-BMR emerges.

\begin{figure}[t!]
      \centering
      \begin{minipage}[b]{1\textwidth}
          \includegraphics[width=0.51\textwidth]{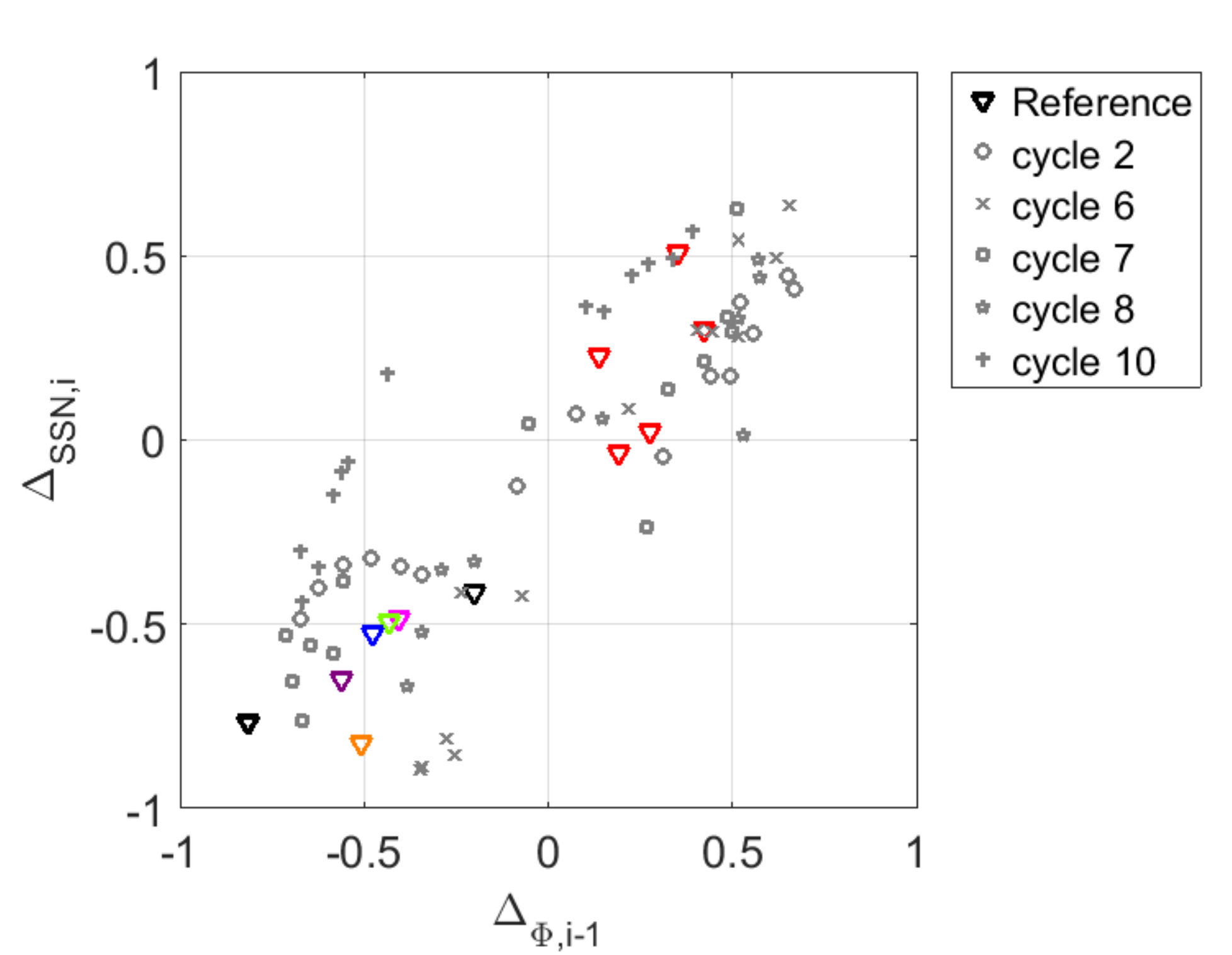}
          \includegraphics[width=0.51\textwidth]{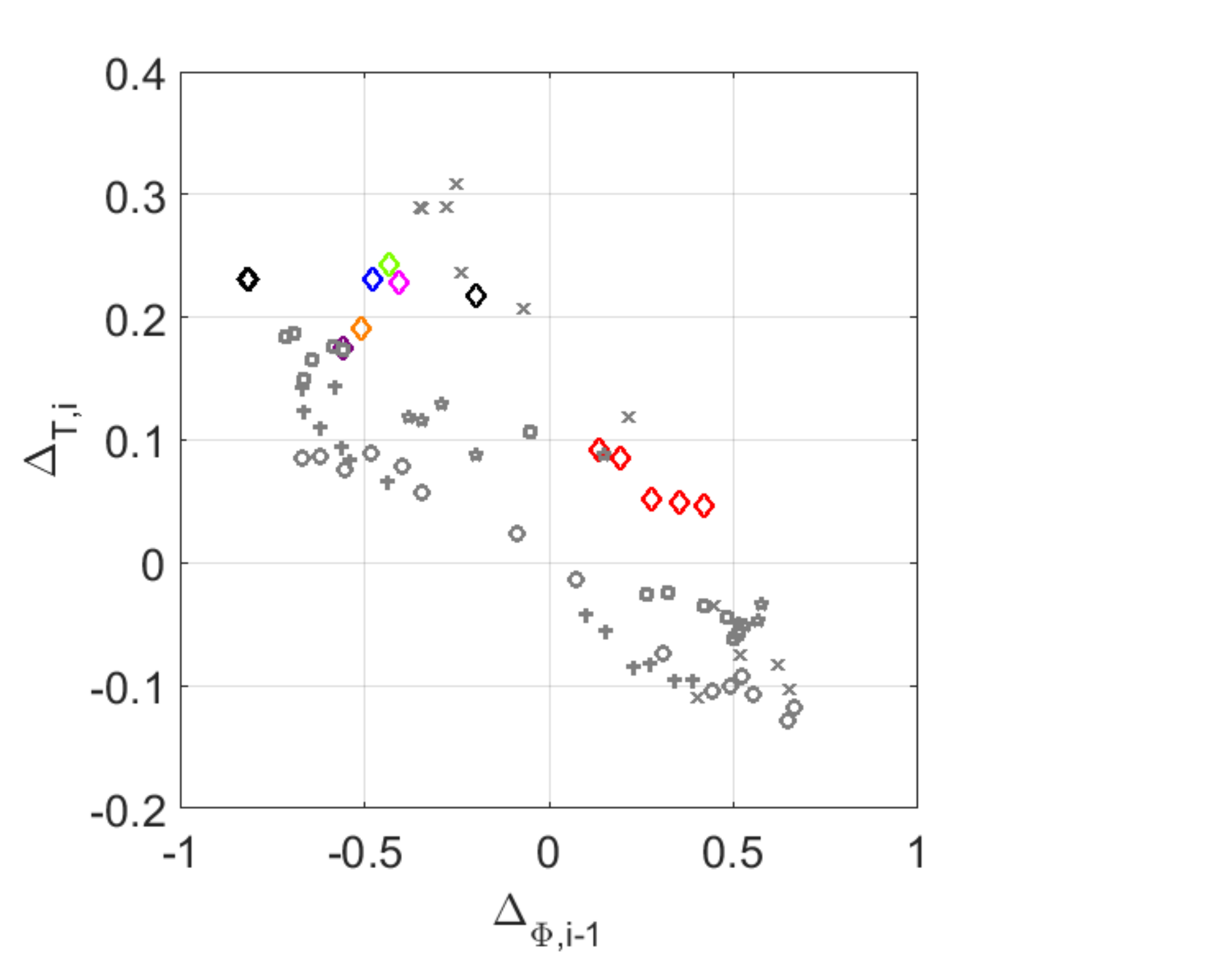}
      \end{minipage}
      \begin{minipage}[b]{1\textwidth}
        \caption{\small{ Correlation plots between the asymmetry of the hemispheric total
     pseudo-SSN (\textbf{left}) and the time lag between North and South (\textbf{right}) in
     pseudo-solar cycle $i$ against the polar cap flux asymmetry during the
     previous cycle. Colored markers correspond to the example shown in Figure \ref{fig:casestudy} and also for the experimental series for the northern hemisphere (red markers). Gray markers show the same series for five more cycles in the same simulation. The correlation coefficients are
     $0.9463$ and $-0.9158$, respectively, and $0.8431$ and $-0.8029$ when using
data from the combined six experiments.  }}\label{fig:asymmetry_seed512}
      \end{minipage}
    \end{figure}

The correlation between the polar cap flux asymmetry triggered by the test-BMR and the asymmetry parameters of the subsequent cycle is plotted in Figure \ref{fig:asymmetry_seed512}. Besides the results for both hemispheres of the reference cycle, we plot results for five more cycles that were studied using the same test-BMR emerging $15^{\circ}$ far from the equator at cycle maximum.
When the BMR is inserted in the northern hemisphere, its tilt and polarity are set to obey Joy's and Hale's law. Considering all the six experimental runs, the correlation coefficient between the polar cap flux asymmetry during the perturbed cycle ($\Delta_{\Phi,i-1}$) and the asymmetry of the number of emerged BMRs during the next cycle ($\Delta_{\mathrm{SSN},i}$) is 0.8431. In the case of the time lag between the hemispheres ($\Delta_{T,i}$) this correlation is $-0.8029$.

Diffusive transport is the only cross-equatorial coupling mechanism operating
in the \cite{Lemerle2017} dynamo model used to carry out the various experiments
described above. The leading member of a BMR emerging $15^\circ$ from the equator
will diffuse to the equator on a timescale $\tau=(\pi R/12)^2/\eta_R\simeq 2\,$yr
for the surface diffusivity value $\eta_R=6\cdot 10^{12}\,$cm$^2\,$s$^{-1}$
used in the SFT module. On the other hand, the internal toroidal field
at 15 degrees will diffuse to the equator on a timescale controlled
by the internal magnetic diffusivity $\eta_t=10^{12}\,$cm$^2\,$s$^{-1}$ of the
FTD module,
leading to a timescale $\tau\simeq 12\,$yr. The first timescale
indicates that a rogue
BMR emerging close to the equator can induce a polar cap asymmetry in the ongoing
cycle, in agreement with the experimental results displayed
on Fig.~\ref{fig:asymmetry_seed512};
while the second timescale reveals that this asymmetry,
once it has built up in the internal toroidal field, can persist over
a full cycle before being diffusively balanced. Periodicity in hemispheric
asymmetry cannot be driven or sustained by such diffusive coupling alone, and
indeed is not observed in our dynamo simulations. Dynamical backreaction on
large-scale flows, namely meridional circulation and differential rotation
would be the most likely candidate mechanism that could lead to such behavior.

\section{Conclusion}

The hemispheric asymmetry triggered by rogue active regions was studied using simulated data of the $2\times2$D solar dynamo model. The flux of a selected test-BMR was changed while its position was fixed to $15^{\circ}$ far from the Equator either
on the northern or southern hemispheres. The emergence epoch was the cycle maximum, while the polarity was set to increase the building up dipole moment. Experimental series were carried out in the case of six simulated cycles of varying amplitudes.

In contrast to the results of \cite{Karak2017} we found strong correlation between the hemispheric asymmetry of the polar cap flux of cycle $i$ and the asymmetry of hemispheric activity levels
during the subsequent cycle $i+1$. The time lag between the hemispheric
lag in the onset of cycle $i+1$ is also strongly correlated to the asymmetry of the
polar cap flux of the preceding cycle. These results can be understood in terms
of diffusive coupling of the magnetic field across the equatorial plane.

Our results thus demonstrate that
the polar cap flux asymmetry at the end of a cycle can be determined
by a single peculiar active region, emerging relatively close to the equator.
This offers an alternate scenario to that suggested by \cite{Hathaway2016},
based on hemispheric variations in the surface meridional flow, which,
in our kinematic dynamo model, remains strictly constant.
In view of the relatively strong
hemispheric asymmetry observed in cycle 24, the unfolding of cycle 25 may
allow to discriminate between these two scenarios.

\section*{Acknowledgements}
M.N.'s research is currently supported by the \'UNKP-18-3 New National Excellence Program of the Ministry of Human Capacities.
P.C. and A.L. are supported through the Discovery Grant Program of the
Natural Sciences and Engineering Research Council of Canada.


\begin{thebibliography}{}

\bibitem[\protect\astroncite{Babcock}{1959}]{Babcock1959}
{Babcock}, H.~D.,
1959, The Sun's Polar Magnetic Field
\newblock {\em ApJ} {\bf 130}, 364-365.

\bibitem[\protect\astroncite{{Belucz} and {Dikpati}}{2013}]{Belucz2013}
{Belucz}, B. and {Dikpati}, M. 2013, Role of Asymmetric Meridional Circulation in Producing North-South Asymmetry in a Solar Cycle Dynamo Model
\newblock {\em ApJ} {\bf 779}, 4, 9 pp.

\bibitem[\protect\astroncite{{Fan}}{2009}]{Fan2009}
{Fan}, Y. 2009, Magnetic Fields in the Solar Convection Zone
\newblock {\em lrsp} {\bf 779}, 6:4.

\bibitem[\protect\astroncite{{Hathaway} and {Upton}}{2016}]{Hathaway2016}
{Hathaway}, D. H. and {Upton}, L. A. 2016, Predicting the amplitude and hemispheric asymmetry of solar cycle 25 with surface flux transport
\newblock {\em JGRA} {\bf 121}, 11, 10,744-10,753

\bibitem[\protect\astroncite{Jiang et al.}{2015}]{Jiang2015}
{Jiang}, J. and {Cameron}, R.~H. and {Sch{\"u}ssler}, M. 2015, The Cause of the Weak Solar Cycle 24
\newblock {\em ApJ} {\bf 808}, L28, 6 pp.

\bibitem[\protect\astroncite{{Karak} and {Miesch}}{2017}]{Karak2017}
{Karak}, B. B. and {Miesch}, M. 2017, Solar Cycle Variability Induced by Tilt Angle Scatter in a Babcock-Leighton Solar Dynamo Model
\newblock {\em ApJ} {\bf 847}, 69, 17 pp.

\bibitem[\protect\astroncite{{Lemerle} and {Charbonneau}}{2017}]{Lemerle2017}
{Lemerle}, A. and {Charbonneau}, P. 2017, A Coupled 2$\times$2D Babcock-Leighton Solar Dynamo Model. II. Reference Dynamo Solutions
\newblock {\em ApJ} {\bf 834}, 133, 18 pp.

\bibitem[\protect\astroncite{{Lemerle} et~al.}{2015}]{Lemerle2015}
{Lemerle}, A., {Charbonneau}, P., and {Carignan-Dugas}, A. 2015, A Coupled $2$D Babcock-Leighton Solar Dynamo Model. I. Surface Magnetic Flux Evolution
\newblock {\em ApJ} {\bf 810}, 78, 18 pp.

\bibitem[\protect\astroncite{{McClintock} and {Norton}}{2013}]{McClintock2013}
B. H. {McClintock}, A. A. {Norton},
 2013, Recovering Joy’s Law as a Function of Solar Cycle, Hemisphere, and Longitude
\newblock {\em SoPh} {\bf 287}, 215-227.

\bibitem[\protect\astroncite{{McIntosh} et~al.}{2013}]{McIntosh2013}
{McIntosh}, S.~W., {Leamon}, R.~J., {Gurman}, J.~B., {Olive}, J.-P., {Cirtain}, J.~W., {Hathaway}, D.~H., {Burkepile}, J., {Miesch}, M., {Markel}, R.~S. and {Sitongia}, L.,
 2013, Hemispheric Asymmetries of Solar Photospheric Magnetism: Radiative, Particulate, and Heliospheric Impacts
\newblock {\em ApJ} {\bf 765}, 146, 17 pp.

\bibitem[\protect\astroncite{{Miesch} and {Dikpati}}{2014}]{Miesch2014}
M. {Miesch}, M. {Dikpati}, 2014, A Three-dimensional Babcock-Leighton Solar Dynamo Model
\newblock {\em ApJ} {\bf 785}, L8, 5 pp.

\bibitem[\protect\astroncite{{Miesch} and {Teweldebirhan}}{2016}]{Miesch2016}
M. {Miesch}, K. {Teweldebirhan}, 2016, A three-dimensional Babcock-Leighton solar dynamo model: Initial results with axisymmetric flows
\newblock {\em AdSpR} {\bf 58}, 1571-1588.

\bibitem[\protect\astroncite{{Mu{\~n}oz-Jaramillo} et~al.}{2012}]{Munoz2012}
{Mu{\~n}oz-Jaramillo}, A., {Sheeley}, N.~R., {Zhang}, J. and {DeLuca}, E.~E.,
 2012, Calibrating 100 Years of Polar Faculae Measurements: Implications for the Evolution of the Heliospheric Magnetic Field
\newblock {\em ApJ} {\bf 753}, 146, 14 pp.

\bibitem[\protect\astroncite{{Nagy} et~al.}{2017}]{Nagy2017}
{Nagy}, M., {Lemerle}, A., {Labonville}, F., {Petrovay}, K., and {Charbonneau}, P.
 2017, The Effect of ``Rogue'' Active Regions on the Solar Cycle
\newblock {\em SoPh} {\bf 292}, 167, 22 pp.

\bibitem[\protect\astroncite{{Norton} et~al.}{2014}]{Norton2014}
{Norton}, A.~A., {Charbonneau}, P., {Passos}, D.,
2014, Hemispheric Coupling: Comparing Dynamo Simulations and Observations
\newblock {\em SSRv} {\bf 186}( 1-4), 251-283.

\bibitem[\protect\astroncite{{Norton} and {Gallagher}}{2010}]{Norton2010}
{Norton}, A.~A. and {Gallagher}, J.~C.,
2010, Solar-Cycle Characteristics Examined in Separate Hemispheres: Phase, Gnevyshev Gap, and Length of Minimum
\newblock {\em SoPh} {\bf 261}, 193-207.

\bibitem[\protect\astroncite{{Shetye} et al.}{2015}]{Shetye2015}
{Shetye}, J., {Tripathi}, D., {Dikpati}, M.,
2015, Observations and Modeling of North-South Asymmetries Using a Flux Transport Dynamo
\newblock {\em ApJ} {\bf 799}, 220, 11 pp.

\bibitem[\protect\astroncite{{Upton} and {Hathaway}}{2018}]{Upton2018}
{Upton}, L.~A. and {Hathaway}, D.~H.,
2018,An Updated Solar Cycle 25 Prediction With AFT: The Modern Minimum
\newblock {\em GRL} {\bf 45}, 6, 8091-8095.

\bibitem[\protect\astroncite{{Waldmeier}}{1955}]{Waldmeier1955}
Waldmeier, M.,
1955, Ergebnisse und Probleme der Sonnenforschung
\newblock {\em Leipzig, Geest \& Portig, 1955.~2.~erweiterte Aufl.} 389 pp.

\bibitem[\protect\astroncite{{Zolotova} et~al.}{2010}]{Zolotova2010}
{Zolotova}, N.~V., {Ponyavin}, D.~I., {Arlt}, R. and {Tuominen}, I.,
2010, Secular variation of hemispheric phase differences in the solar cycle
\newblock {\em AN} {\bf 331}, 765-771.

\end{thebibliography}
\end{document}